\documentclass[aps,prstab,twocolumn,amsmath,amssymb,groupedaddress]{revtex4-2}

\usepackage{siunitx}
\usepackage{placeins}
\usepackage{float}
\usepackage[colorlinks=true]{hyperref}
\hypersetup{ 
    citecolor=black,%
    filecolor=black,%
    linkcolor=black,%
    urlcolor=black 
}
\usepackage{graphicx}
\usepackage{dcolumn}
\usepackage{bm}
\usepackage{color}

\begin{document}

\title{A final focus system for injection into a laser plasma accelerator at the ARES linac}

\author{S.~Yamin}
\email{sumera.yamin@desy.de}
\affiliation{DESY, Notkestrasse 85, 22607 Hamburg, Germany. Also at University of Hamburg, 20148, Hamburg, Germany}
\author{R.~Assmann}
\affiliation{DESY, Notkestrasse 85, 22607 Hamburg, Germany}
\author{F.~Burkart}
\affiliation{DESY, Notkestrasse 85, 22607 Hamburg, Germany}
\author{A.~Ferran Pousa}
\affiliation{DESY, Notkestrasse 85, 22607 Hamburg, Germany}
\author{W.~Hillert}
\affiliation{University of Hamburg, 20148, Hamburg, Germany}
\author{F.~Lemery}
\affiliation{DESY, Notkestrasse 85, 22607 Hamburg, Germany}
\author{B.~Marchetti}
\affiliation{DESY, Notkestrasse 85, 22607 Hamburg, Germany}
\author{E.~Panofski}
\affiliation{DESY, Notkestrasse 85, 22607 Hamburg, Germany}

\date{\today}

\begin{abstract}
ARES (Accelerator Research Experiment at SINBAD) is a linear accelerator at the SINBAD (Short INnovative Bunches and Accelerators at DESY) facility at DESY. ARES was designed to combine reproducible beams from conventional RF-based accelerator technology with novel but still experimental acceleration techniques. It aims to produce high brightness ultra-short electron bunches in the range of sub fs to few fs, at a beam energy of 100-150 MeV, suitable for injection into novel acceleration experiments like Dielectric Laser Acceleration (DLA) and Laser driven Plasma Acceleration (LPA). This paper reports the conceptual design and simulations of a final focus system for injecting into a LPA experiment at ARES, including permanent magnetic quadrupoles (PMQ), sufficient longitudinal space for collinear laser and electron transport, space for required diagnostics and a LPA setup. Space-charge effects play a significant role and are included. Simulation results on the focusing of the ARES electron bunches into and their transport through the laser-driven plasma are presented. The effects of several errors have been simulated and are reported.
\end{abstract}

\maketitle

\section{Introduction}\label{sec:Introduction}
SINBAD acronym for Short INnovative Bunches and Accelerators at DESY, is an accelerator R\&D facility at DESY on its Hamburg site \cite{Assmann:IPAC2014-TUPME047}. It includes research in the field of ultrashort electron bunches and will host multiple independent experiments \cite{Dorda:NIMA2018SINBAD} on laser driven advanced high gradient acceleration techniques such as Dielectric Laser Acceleration (DLA) \cite{Mayet:2018hj}, Laser Plasma Acceleration (LPA) \cite{Marchetti:2018AppSciences} and THz-driven acceleration in the AXSIS project \cite{Kartner:NIMA2016AXSIS}. The ARES (Accelerator Research Experiment at SINBAD) linear accelerator (linac) at SINBAD  is based on conventional S-band technology with a photo-injector gun \cite{Panofski:IPAC2019-MOPTS026}\cite{MarchettiIOP2020SINBAD_ARES}. It is designed to provide ultrashort high brightness electron beams for injection into novel accelerators. The FWHM length of electron bunches is expected to reach a few fs and potentially sub fs values. The electron gun relies on a conventional radio frequency (RF) accelerator technology for producing the electron bunches. This has several advantages. The ARES linac on one hand will allow advancing R\&D on the “conventional” production of high brightness ultra-short electron bunches. On the other hand, the well-characterized bunches can be used to explore compact novel accelerators, characterized by accelerating fields with short wavelengths and therefore require the injection of short “acceleration buckets”. The ARES bunches have been designed to constitute excellent probes to measure the energy gain and the quality of the acceleration. 

At ARES, the electron bunches can be compressed either via velocity bunching or by using a magnetic bunch compressor or by using a hybrid scheme to achieve the desired bunch lengths in the range of sub fs to few fs \cite{Marchetti:IPAC2015-TUPWA030}\cite{Zhu:2016PRABsubfselwithsubfsbuncharrivaltime}. The characterization of such ultra-short bunches is a research field in itself and ARES will also serve as a test bench for novel diagnostic devices in the low to medium charge range of up to 30 pC and with sub-fs to few fs bunch length \cite{Merz:2019EAACStridenas}\cite{Marx:2019Nature}. All these features contribute in making ARES a promising candidate where LPA experiments with external injection could be performed. This potential was explored as a part of a PhD project.

LPAs can provide an accelerating gradient in the range of $\approx$100 GV/m, which is several orders of magnitude higher than what can be achieved with conventional RF technology. This reduces the size of the acceleration channel from meters to millimeters, to achieve GeV energy range, and hence offers the possibility of compact and cost-effective accelerators. In a LPA, strong laser pulses propagating in a plasma generate charge separation through the excitation of wakefields, inducing strong electric fields. Since the LPA concept was first introduced \cite{Tajima:PRL1979LWFA}, it has been a subject of extensive studies with significant progress in recent years \cite{Leemans:2006NAtureGeVelectronbeam}\cite{Litos:2016IOP9GeVenergygain_beamdriven}. Many important milestones have been demonstrated such as achieving GeV energies in only cm scale \cite{Gonsalves:2019PRL8GeV}. Today’s focus for these devices is to reach beam stability, similar to established RF accelerators, by deploying feedback control systems. Recently another milestone has been successfully demonstrated at DESY with 24 hours of stable operation of laser plasma acceleration \cite{Maier:2020PRXstability}. The external injection LPA experiment studied for ARES, investigated a possible step towards usable LPA from a combination of reproducible conventional RF-based accelerator technology with the high gradient fields from plasma wakefields \cite{Grebenyuk:IPAC2014-TUPME064}. The RF-based technology allows for a precise manipulation of the phase space of the electron bunches before entering the plasma hence, providing independent control and quality adjustments as well as optimization of the plasma experiment. The beam quality in novel accelerators depends on the detailed parameters and quality of the injected beam e.g. bunch shape, bunch length, emittance, arrival time stability and beam energy. ARES provides the option of widely tunable working points (WP) and bunch shapes for external injection LPA accelerator R\&D. External injection of electron beam into an LPA, however, has its own technical challenges \cite{MarchettiIOP2020SINBAD_ARES}. The synchronization of laser and electron beam can be the most crucial aspect of this experiment.

The present status of ARES is shortly summarized. Its 5 MeV RF gun and linac has been commissioned \cite{Marchetti:IPAC2018-TUPMF086} and first electrons have been produced at the end of 2019 [19]. The installation of the subsequent experimental chamber and diagnostic beam line is finished \cite{Burkart:IPAC2019-MOPTS014}. The preparations for a DLA experiment are ongoing and simulation studies have been performed for a potential LPA experiment \cite{Svystun:IPAC2019LWFA_ARES}.
In this paper, we present design studies for a final focus system that could fulfill the requirements of external injection of ARES bunches into a LPA setup. In the following sections, the layout of the ARES linac and the experimental area for the simulated LPA experiment are described, followed by the requirements for beam matching into the plasma cell. The results for design and optimization of the final focus system and for electron beam tracking through the lattice are presented. Tolerance studies for the final focus system are presented and discussed. 

\section{Layout of ARES linac} \label{sec:Layout}

\begin{figure*}[htbp]
	\centering
	\includegraphics[width=0.99\textwidth]{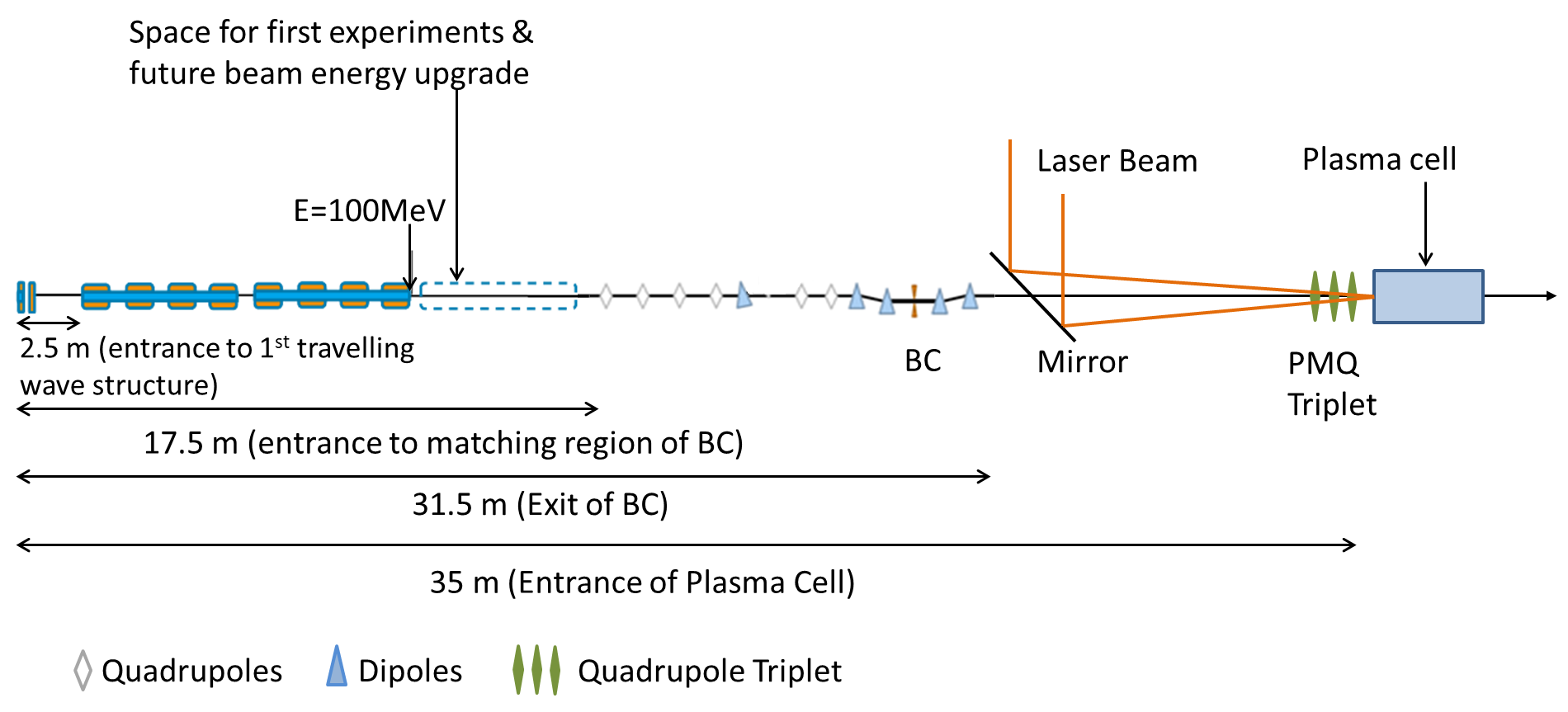}
	\caption{Schematic of ARES}
	\label{fig:SchematicARES}
\end{figure*}

A schematic overview of the ARES linac with a potential LPA acceleration experiment is shown in Fig.~\ref{fig:SchematicARES}. The main beam parameters are summarized in Table 1. The ARES linac consists of a 5 MeV RF gun followed by two travelling wave structures (TWS) with space reserved for a third travelling wave structure for a possible future energy upgrade \cite{MarchettiIOP2020SINBAD_ARES}. This space will be temporarily used as first experimental area (EA). This is followed by a matching section into a magnetic bunch compressor (BC). At the exit of the BC the electron bunches have a duration of a few fs or below \cite{Lemery:IPAC2019-MOPTS025Bunchcompressor}. The BC is followed by a drift space that provides space for the laser incoupling. Further downstream, we have a high energy diagnostic beamline followed by the matching optics for the second experimental area. This area could host LPA experiment and could combine the electron beam from ARES with high power laser pulses from the high repetition high power laser KALDERA currently under construction at DESY \cite{KALDERA}. The matching scheme is designed for the low charge WP of the ARES linac that will provide 0.8 pC electron bunches with smallest arrival time jitter of about $\approx$10 fs rms. It uses a pure magnetic compression scheme and is designed for ultimately sub fs bunch length. The WP has been discussed in detail in \cite{Yamin:IOP2020PMQ}. The electron bunches from this setup provide a time resolution in sub fs range and can serve as probe particles for plasma wakefields or fields in dielectric structures. 

In this paper, we present the results for the higher charge 10 pC WP of the ARES linac. It features a longer bunch length of 10 fs FWHM or 4.3 fs rms. The peak current in this case is 1 kA, approaching the requirements of several use cases \cite{Pousa:PRL2019Multistagedechirping}. It is noted that this WP is similar to a WP studied at the BELLA facility in LBNL for which a broad energy spread electron bunch of $\approx$ 100-200 MeV energy, 6.25 pC charge and 3.3 fs rms length was used for injection into a second stage LPA \cite{Tilborg:AIP2017FELbyPlasmaacelerators}. 

\begin{table}[!tbp]
	\centering
	\caption{Parameters for the RF systems and the electron beam at the ARES linac}
	
	\begin{tabular}{lc}
		\hline
		\textbf{Parameter}	         &\textbf{Values}     \\
		\hline
		RF Frequency                 & 2.998 GHz        \\
		Repetition rate              & 50 Hz         \\
		Beam Energy 
		  & 100 MeV 
		(155 MeV on crest)        \\
		Upgraded beam energy 
		& 150 MeV 
		(230 MeV on crest)\\
		Bunch Charge      & 0.5-30  pC        \\
		Bunch Length        & sub to few fs           \\
		 Arrival Time Jitter stability      & 10fs to few tens of fs    \\
		\hline
	\end{tabular}

	\label{tab:ARESbasicParameters}
\end{table}

\section{Technical constraints for the matching beam line} \label{sec:Constraints_matchingbeamline}

In the external injection LPA experiment, the laser beam and the electron beam are collinearly injected into the plasma channel. The collinear electron and laser beam lines, along with required beam diagnostic elements, introduce strict constraints on both the transverse and longitudinal dimensions for the design of the final focusing system. The system must be compact and must provide gradients high enough to tightly focus the beam into the plasma cell. Those constraints led to choosing a PMQ triplet as final focusing system.

\begin{figure}[htbp]
	\centering
	\includegraphics[width=\columnwidth]{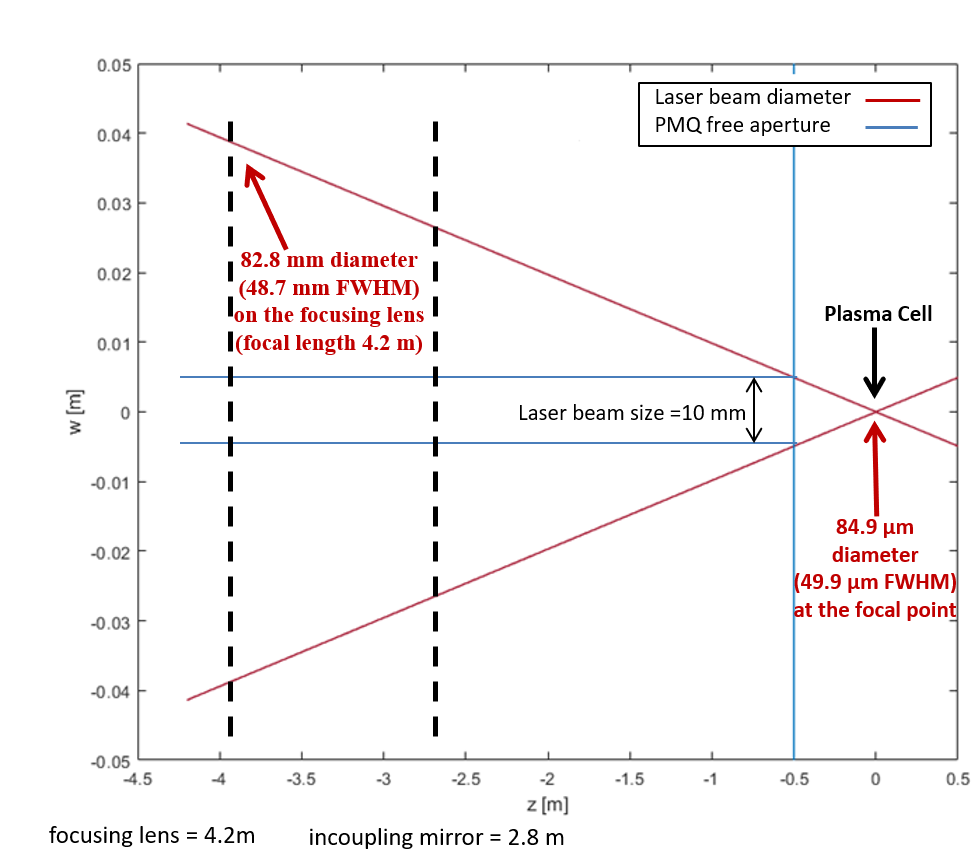}
	\caption{Evolution of the laser beam along the direction of beam propagation. The origin here is set at the focal point of the laser beam which is the entrance position of the plasma cell. A Gaussian laser beam is assumed.}
	\label{fig:LaserEvolution}
\end{figure}

The key parameters for the laser system are presented in table~\ref{tab:LaserParameters}. The schematic of the laser beam at its waist and the parameter definitions are shown in Fig.~\ref*{fig:LaserEvolution}. For this study a Gaussian laser beam is assumed. Since the divergence of the laser beam depends on the beam waist w\textsubscript{o}, the design of the laser beamline and hence in turn the space constraints for our final focus system are strongly dependent on the laser parameters.

\begin{table}[htbp]
	\centering
	\caption{Key parameters for the Laser setup}
	\begin{tabular}{lcc}
		\hline
		\textbf{Parameters} & \textbf{Units} & \textbf{Values}\\
		\hline
		Wavelength $\lambda_o$ & $\mu$m & 0.8 \\
		Vector potential $\text{a}_\text{o}$ & 1.4\\
		Beam waist $w_o$ & $ \mu$m&42.4  \\
		Pulse Energy (E) & J & 3 \\
		Peak Power (P) & TW & $ \sim $100\\
		Pulse length ($\Delta$t\textsubscript{FWHM}) &fs& 30\\
		\hline
	\end{tabular}
	\label{tab:LaserParameters}
\end{table}

The technical design considerations for the PMQ triplet have been discussed in detail in \cite{Yamin:IPAC2019-PMQ_MOPGW027}. The required mirror is housed in a beam pipe having diameter of 10 cm. The beam pipe size is chosen to accommodate the mirror dimensions required for focusing the 100 TW peak power laser beam. A hole in the mirror allows the electron beam transmission and, collinearly to the laser beam, the electron beam enters the plasma cell. At the focal point, the laser beam has a design waist of 42.4 $\mu$m. For the external injection experiment, the laser beam also has to pass through the PMQ triplet to reach the plasma cell. Hence the free aperture, defined by the distance of the magnetic pole tips of the PMQ should be bigger than the laser spot size at the position of quadrupoles. Considering the laser beam evolution from Fig.~\ref*{fig:LaserEvolution}, the aperture of PMQ is chosen to be 10 mm. The laser parameters and the laser beam line design dictate the limits for the physical dimensions of the PMQ inner and outer radii, the length of the total triplet and also set constraints to the positions of the focusing magnet. The distance between the exit of the BC and the entrance of the plasma cell is $\approx$3.3 m to account for the space required for diagnostics, the vacuum system and for focusing the laser beam. Since the divergence of the laser beam depends on the beam waist w\textsubscript{o}, the design of the laser beamline and hence in turn space constraints for our final focus system have a strong dependence on the laser parameters. 

\section{Matching criterion} \label{Matching criterion}

For preserving the beam emittance, it is required to match the externally injected electron beam to the focusing fields of the plasma accelerator. The matching conditions depend on the focusing strength of the plasma channel, essentially given by the plasma density \cite{Li:2019PRABpresEmitMatchInMatchOut}. For the LPA experiment studied, the matched Twiss parameter is $\beta_{x,y}$ $\approx$ 1 mm in case of a step function of the longitudinal plasma density profile. The density of the plasma channel is $10^{17}$ $cm^{-3}$. This requirement was relaxed by proper shaping of the longitudinal plasma density profile (upramps) as discussed in \cite{Xu:2016PhSpmatchingusingLongPlasmaProfiles} and \cite{Floettmann:PRAB2014AdiabaticMatching}. The longitudinal plasma density profile given by Eq.~\ref{eq:PlasmaDensityProfileCase2} was implemented to match the electron beam into the plasma focusing field and thus to control the emittance growth of the beam in the plasma section:

\begin{equation}
\label{eq:PlasmaDensityProfileCase2}
n_{p,r}=\frac{n_{p,o}}{ \left( 1+\frac{~z}{L_{r}} \right) ^{2}} \
\end{equation}

where \textit{n\textsubscript{p,r}} is the plasma density, \textit{n\textsubscript{p,o}} is the density of the plateau, \textit{z} is the distance to the plateau and \textit{L\textsubscript{r}} is the length parameter which determines how fast the density decreases with z. Such a ramp has been found to have good performance as illustrated in \cite{Xu:2016PhSpmatchingusingLongPlasmaProfiles}. The optimization of the plasma ramps was done via FBPIC simulation scans in which different values for \textit{L\textsubscript{r}} were simulated and optimized. For the present case, \textit{L\textsubscript{r}} = 2 mm was used. The calculation of the necessary Twiss parameters at the plasma entrance were obtained by back-tracking a matched beam at the plateau entrance, through the ramp. The back-tracking in this case implies that the beam was propagated through a downramp identical to the upramp. The approach was to calculate the beta function needed at the beginning of the plateau using the focusing fields there. Then a Gaussian bunch was generated with this beta and alpha = 0 and then tracked, including the laser driver, through a plasma down ramp which has the same shape as the up ramp. This is equivalent to back propagating the bunch through the plasma. A similar approach has been used in \cite{Dornmair:2015EmitConsbytailoredfocusingprofiles}. The required matched twiss parameters for a laser spot size of 40 $\mu$m are then  \textit{$ \beta $ \textsubscript{x,y}} = 11.8 cm and \textit{$ \alpha $ \textsubscript{x,y}} = 4.4. Throughout the article, the entrance of plasma cell refers to the start of plasma ramps for which this matching criterion is defined. 

\section{Design study for the matching beamline}\label{sec:design_study}

The beam dynamics simulations for the matching beamline after the BC have been performed using Elegant (without space charge) \cite{elegantAFlexible:DccEjrWC}  and then ASTRA \cite{ASTRAASpaceChar:LTSRiAsm} to optimize and include the effects of 3D space charge (SC). The optimization parameters for the PMQ triplet are lengths, strengths and distances between the quadrupoles. Based on the laser beamline layout, as shown in Fig.~\ref{fig:LaserEvolution}, the constraints for the final focus system were the quadrupole aperture, the outer diameter corresponding to size of the quadrupoles and the focal length. The focal length must take into account space for diagnostic screens allowing laser and electron beam profiling between the last magnet of the triplet and the plasma cell. We set the origin of our simulations at the exit of the last magnet of the BC and the beamline is simulated from this point until the entrance of the plasma cell corresponding to the start of plasma upramp as defined by the matching criterion in Section~\ref{Matching criterion}. 

\begin{figure*}[htbp]
	\centering
	\includegraphics[width=0.99\textwidth]{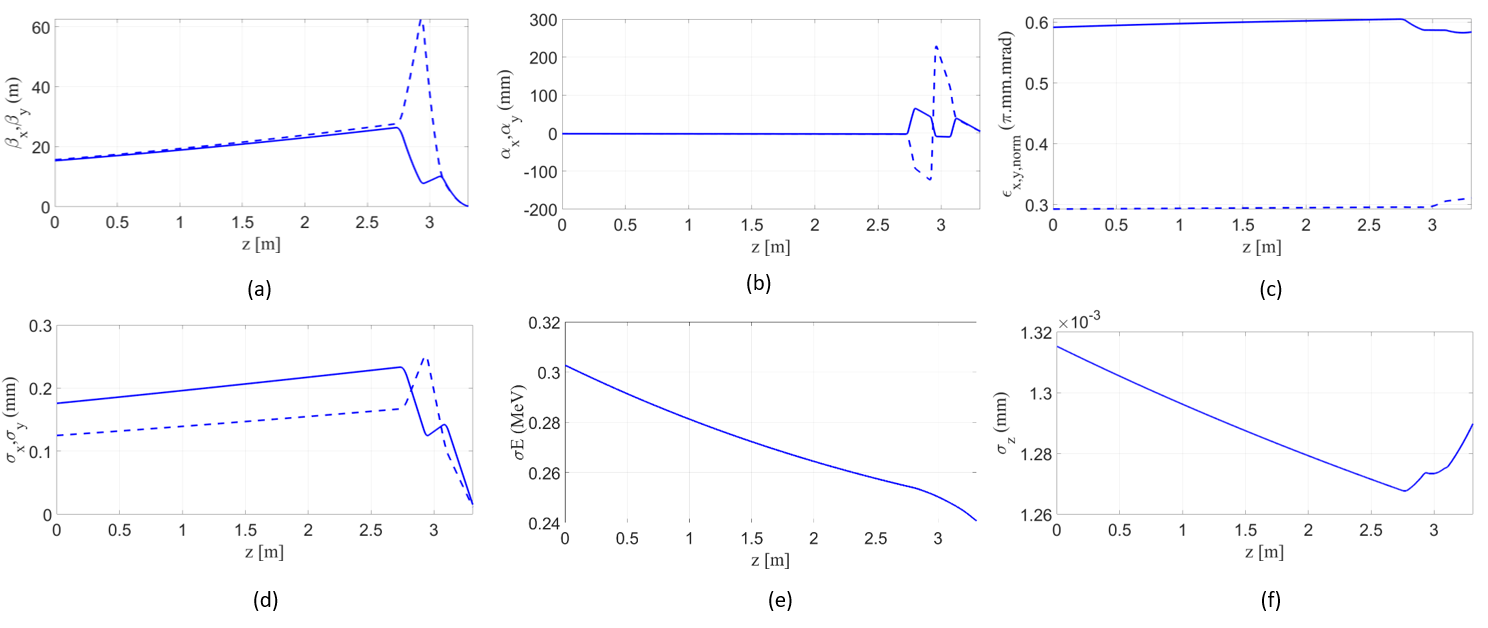}
	\caption{Evolution of the electron beam parameters along the beam line from the exit of the BC until the entrance of the plasma cell. The origin here is set at the exit of the last magnet of the BC. The parameters shown are the Twiss parameters in (a) $\beta_{x,y}$ and in (b) $\alpha_{x,y}$, in (c) the normalized transverse emittance $\varepsilon_{x,y}$, in (d) the rms transverse beam sizes $\sigma_{x,y}$, in (e) the energy spread $\sigma_{E}$ and in (f) the bunch length $\sigma_{z}$}
	\label{fig:BeamEvolution_3TWS_150MeV}
\end{figure*}

\begin{figure*}[htbp]
	\centering
	\includegraphics[width=0.99\textwidth]{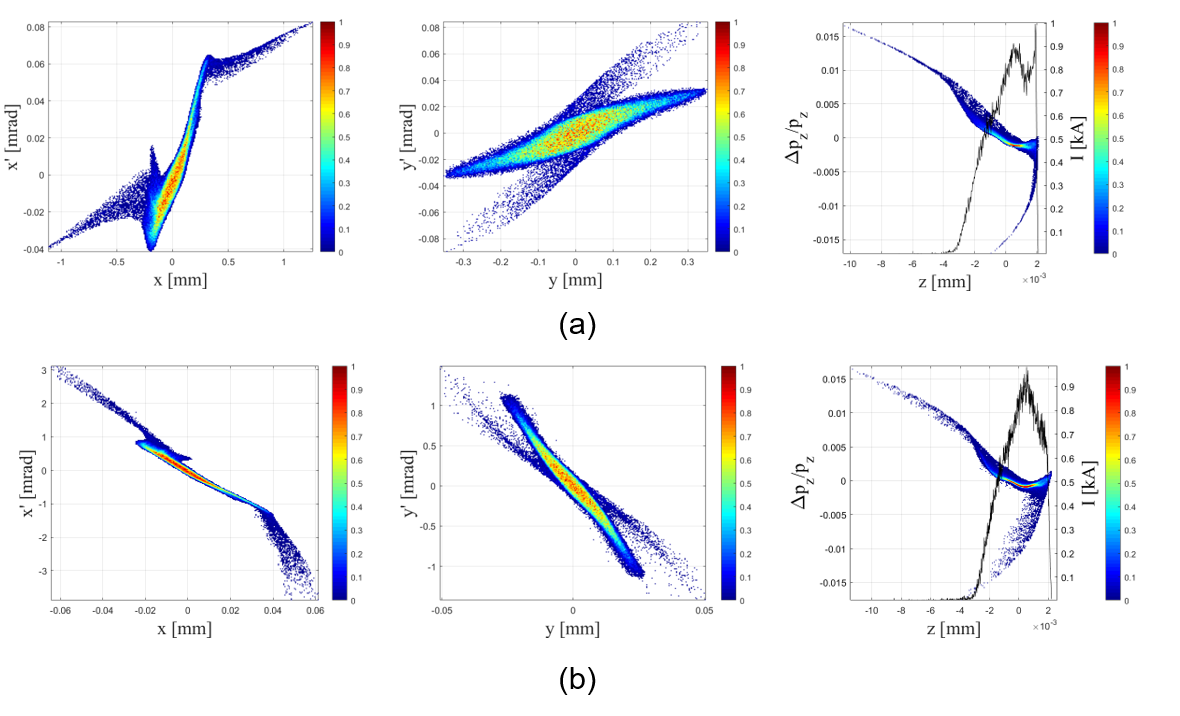}
	\caption{Evolution of the transverse and longitudinal phase spaces at (a) the BC exit and matched beam at (b) the entrance of the plasma cell. Color scales indicate normalized electron density.}
	\label{fig:PhaseSpaces_3TWS_150MeV}
\end{figure*}

The key parameters for the working points at the BC exit and at the entrance of the plasma cell are summarized in Table~\ref{tab:Beamparameters3TW}. The full beam distribution is included in the calculation of statistical parameters. The three quadrupoles of the triplet have lengths of 0.055 m, 0.033 m and 0.040 m with strengths of  50 m\textsuperscript{-2}, -180 m\textsuperscript{-2} and 130 m\textsuperscript{-2} respectively. The distance between the last magnet and the plasma cell is 0.25 m which is sufficient to place screens for laser and electron beam profiling. Fig.~\ref{fig:BeamEvolution_3TWS_150MeV} shows the evolution of the electron beam parameters over the drift space and through the PMQ triplet from the BC to the plasma cell. The transverse and longitudinal phase spaces of the electron beam at the BC exit and at the entrance of plasma cell are shown in Fig.~\ref{fig:PhaseSpaces_3TWS_150MeV}. From Fig.~\ref{fig:BeamEvolution_3TWS_150MeV} and \ref{fig:PhaseSpaces_3TWS_150MeV} it is evident that at the start of plasma ramp, the electron beam is well matched with preserved transverse emittance and bunch length shorter than the accelerating wavelength in the plasma. The transverse phase space distributions show that the beam is symmetric in both planes as required by LPA. The PMQ triplet also ensures to maintain the current profile from the longitudinal phase space with $\approx$1 kA of peak current.
 
Detailed simulation studies on beam tracking in the experimental beamline through the PMQ triplet show that the optimized settings for the PMQ triplet provide a good quality working point. The electron beam can be transported and matched into the plasma cell with the PMQ triplet while preserving the beam quality. The transverse and longitudinal phase space distributions show good beam profiles for injecting into LPA. The symmetric beam distributions with a peak current of 1 kA would be sufficient for efficient beam loading in the plasma cell.

\begin{table}[htbp]
	\centering
	\caption{Bunch Parameters at the BC exit (I) and of matched beam at the entrance of plasma cell (II)}
	\begin{tabular}{lcc}
		\hline
		\textbf{Parameters}	& \textbf{I} &\textbf{II}   \\
		\hline
		Energy (MeV)                     & 150      & 150  \\
		Bunch Charge (pC)                & 10       & 10  \\
		Bunch length FWHM/rms (fs)        &11.8/4.38 & 11.9/4.30 \\
		$\varepsilon_x$ ($\pi$.mm.mrad)  &0.59      &0.57   \\
		$\varepsilon_y$ ($\pi$.mm.mrad)  & 0.29     &0.31   \\
		$\beta_{x}$ (cm)                 &15.3e2    &11.8  \\
		$\beta_{y}$ (cm)                 & 47.22e2  & 11.9 \\
		$\alpha_{x}$	                 &-1.7      &4.5 \\
		$\alpha_{y}$	                 &-1.7      &4.4  \\
		$\sigma_{x}$                  	 &175.8     &15.2   \\
		$\sigma_{y}$	                 &124.9     &11.2   \\
		RMS energy spread (\%)           &0.2       &0.16  \\
		Peak Current (kA)                &1         &1   \\
		\hline
	\end{tabular}
	\label{tab:Beamparameters3TW}
\end{table}

\section{Error analysis of the PMQ triplet and the incoming beam}\label{sec:ErrorAnalysis}

The experimental conditions may vary from the ideal design parameters discussed in the previous sections due to temporal and spatial jitter in the beam. Moreover, there are additional sources of errors in the quadrupoles, arising during the manufacturing and installation of the final focus system. For example, offsets between the magnetic field center of the quadrupoles and the ideal beamline can lead to emittance growth and betatron oscillations in the plasma. A basic tolerance study for the final focus system was performed to estimate the effects of several error sources.
 
We consider two sources of errors in our system. One set of errors can arise from spatial and temporal jitter in the electron beam. The other set of errors is introduced from positioning errors in the quadrupole triplet. In simulation parameters in the beam distribution after the bunch compressor were varied and analyzed. In addition, perturbations from transverse offsets x\textsubscript{off} and y\textsubscript{off} in the quadrupole triplets, from rotation errors x\textsubscript{rot} in the x-z plane, from rotation errors y\textsubscript{rot} in the y-z plane, from rotation errors z\textsubscript{rot} in the x-y plane, from longitudinal offset errors z\textsubscript{off} and from errors in the focusing strength K were included. The quadrupole parameters under study were varied according to Eq.~\ref{eq:TolRefRandLatticeDef}

\begin{equation}
\label{eq:TolRefRandLatticeDef}
A=A_{ref} \pm A_{off} \left(  \pm tol \right) 
\end{equation}

where \textit{A} represents an input parameter, \textit{A\textsubscript{ref}} is the design parameter of the quadrupole and \textit{A\textsubscript{off} ($\pm$ tol)} is the variation of the design parameter within the given tolerances range. All quadrupoles were assigned the same error as one would realistically expect if errors arise from a common power supply or a common support girder. For example, mounting errors between single elements on a common girder are usually much better corrected than offsets or drifts of the whole girder in the tunnel. 

For the tolerance studies for jitter in the input beam, the beam properties were perturbed according to Eq.~\ref{eq:TolRefRandBeamDef}. Included were transverse offsets in electron bunch position in both transverse planes (x and y), errors in bunch size in both transverse planes, the beam momentum and divergence, bunch charge, bunch length as well as longitudinal offset in z, all defined at the exit of the bunch compressor: 

\begin{equation}
\label{eq:TolRefRandBeamDef}
\Delta b_i=b_{i,ref} \pm b_{off}\left( \pm tol \right)  
\end{equation}

where \textit{b\textsubscript{i,ref}} is the reference beam parameter at the exit of BC given in Table~\ref{tab:Beamparameters3TW} and \textit{b\textsubscript{off}} is the error in the beam parameter at the same location. 

The observables in simulation include transverse normalized emittance, transverse beam sizes, beam divergence in both planes, Twiss parameters, bunch length and energy spread at the entrance of the plasma cell. 

In order to assess the robustness of the PMQ triplet solution, 10,000 cases were simulated for two cases.

Mismatch case 1: Errors are randomly assigned. This includes $\pm$10 $\mu$m errors in bunch position at the BC exit and 10\% errors in beam size, bunch length and bunch charge. For the quadrupole triplet transverse and longitudinal offsets were varied within ± 10 $\mu$m, rotational offsets within $\pm$ 10 $\mu$rad and the strength were varied by $\pm$ 10\% of the design value. The minimum tolerance range of 10 $\pm$m was chosen according to the precision range of hexapods which can be used for the positioning of the quadrupole triplet \cite{SmaractHexapods}.

Mismatch case 2: A same set of simulations were carried out in which the tolerance range for transverse and rotational offset of quadrupole was 100 $\mu$m and 100 $\mu$rad respectively.

\begin{figure*}[htbp]
	\centering
	\includegraphics[width=0.99\textwidth]{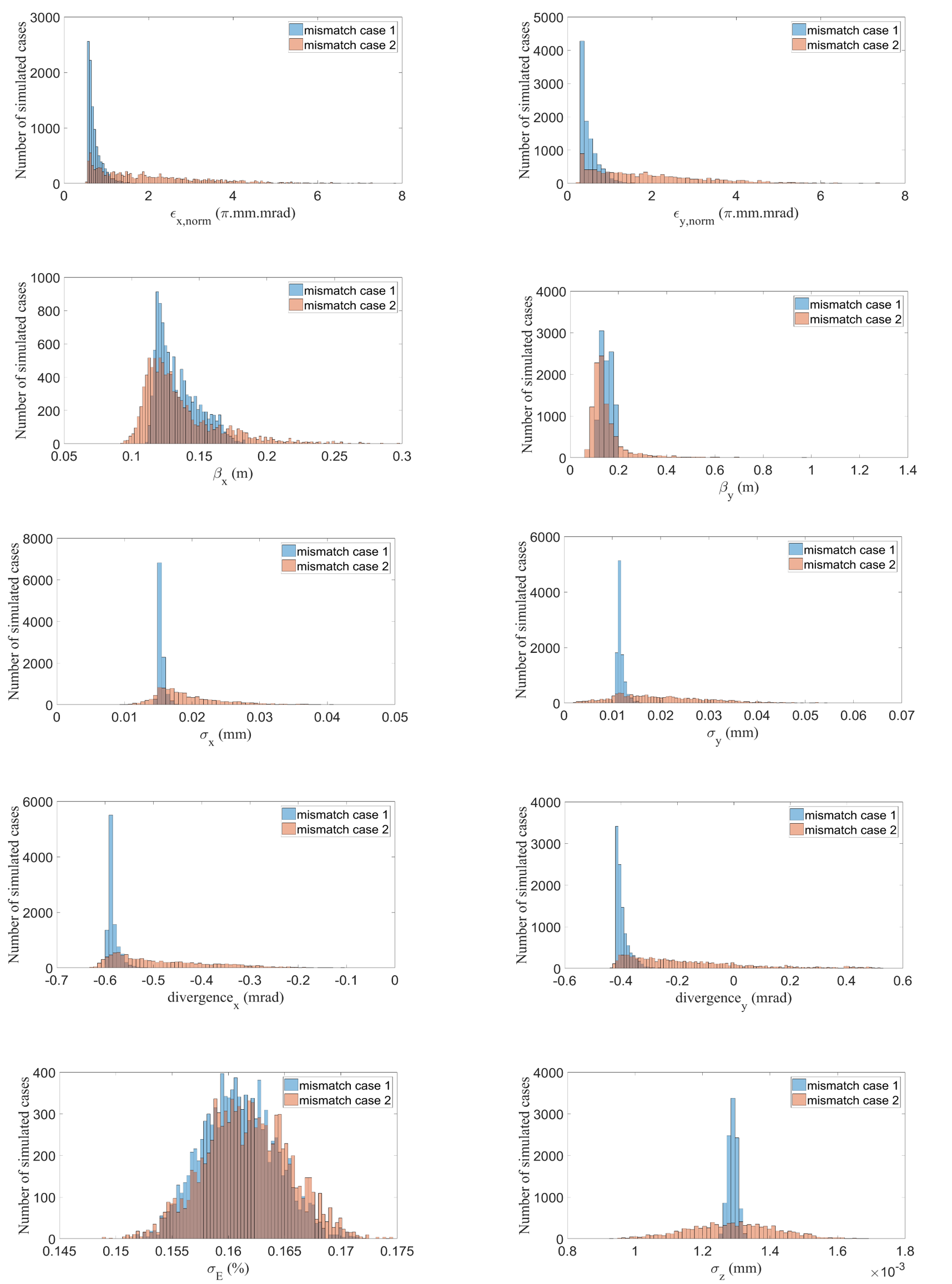}
	\caption{Variation in observed beam parameters at the plasma entrance for the two error scenarios defined in the text (mismatch case 1 and 2) and each 10,000 random cases simulated}
	\label{fig:Histograms150}
\end{figure*}

The position for observing the final beam parameters in both cases was fixed at the matched case position of Table~\ref{tab:Beamparameters3TW}, which is the entrance of plasma cell as explained in section~\ref*{Matching criterion}. Fig.~\ref{fig:Histograms150} summarizes the simulated beam parameters at the plasma entrance, as obtained in the two error cases. From Fig.~\ref{fig:Histograms150}, it can be concluded that the mismatch case 1 is an acceptable scenario, since the beam with this set of variations still enters the plasma cell well matched in the transverse plane with only a small increase in emittance. It is worthwhile to note again that this also includes the variation in input beam parameters. It can be safely inferred that the PMQ triplet design is robust and can be used for matching the electron beam to the plasma channel under less than ideal experimental conditions as well.

\begin{figure}[htbp]
	\centering
	\includegraphics[width=\columnwidth]{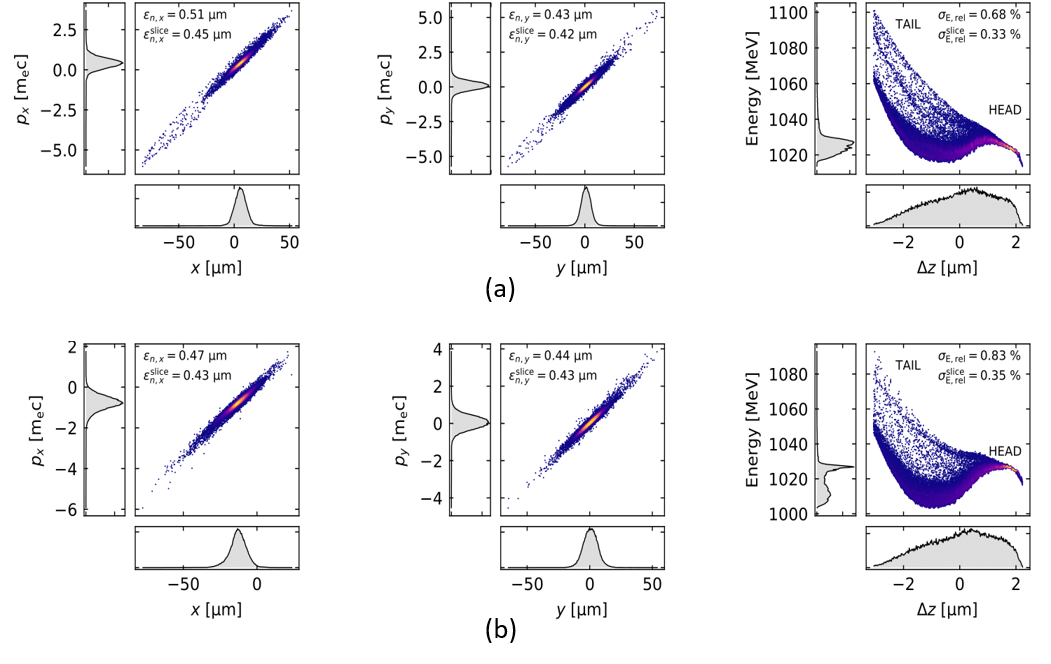}
	\caption{Simulated evolution of beam parameters through a LPA plasma cell in the ARES linac for the matched beam and the mismatched beam considered in this study}
	\label{fig:phasespaces_combined_inplasma_150MeV}
\end{figure}

\section{Simulated acceleration through a plasma cell}

\begin{figure}[!tbp]
	\centering
	\includegraphics[width=\columnwidth]{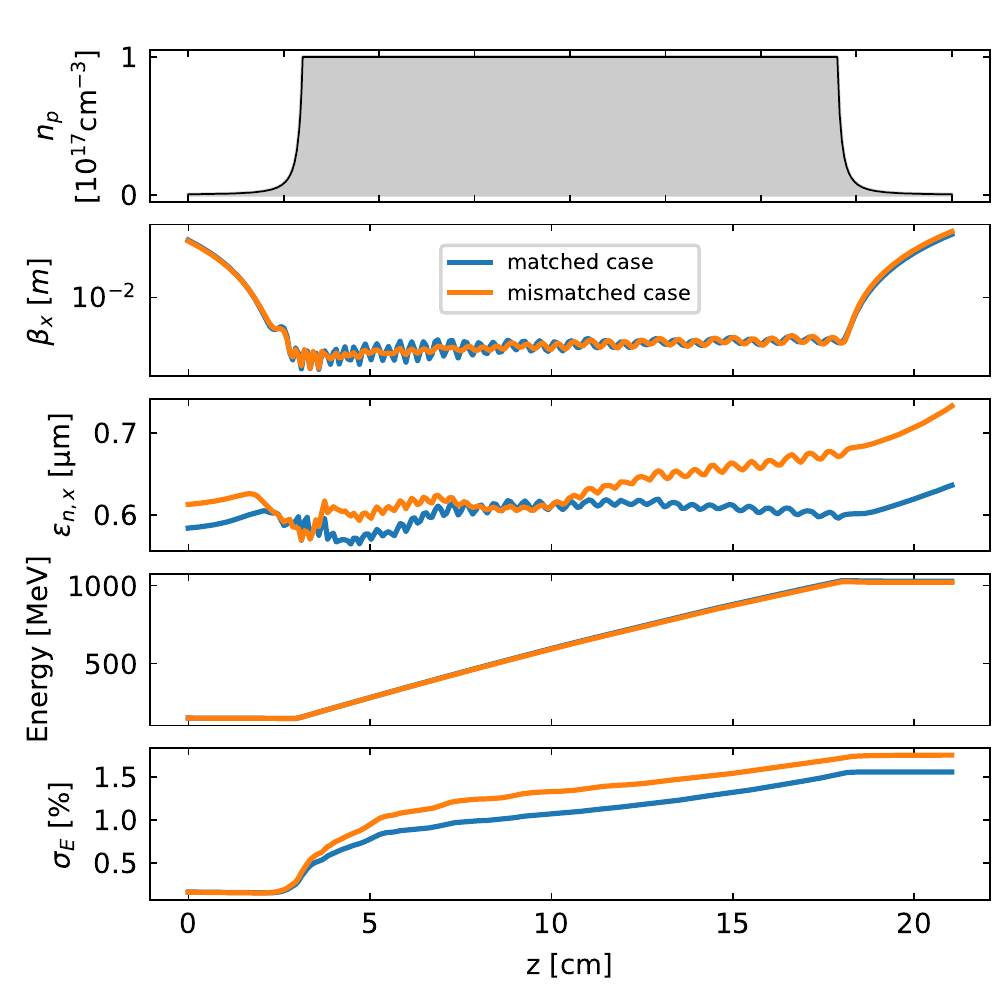}
	\caption{Simulated evolution of beam parameters through a LPA plasma cell in the ARES linac for the matched beam and the mismatched beam considered in this study}
	\label{fig:beamevolution_inplasma_150MeV}
\end{figure}

The simulated beam matched to the plasma entrance (reference case from Section~\ref{sec:design_study}) was further tracked through a laser-driven plasma cell (LPA) using the FBPIC code \cite{Remi:2016FBPIC}. A plasma cell model, with density of $10^{17}$ $cm^{-3}$ is used to accelerate the beam. Fig.~\ref{fig:beamevolution_inplasma_150MeV} shows the simulated beam evolution through the plasma cell. The density profile and the evolution of the Twiss parameter $\beta_{x}$, the normalized emittance $\varepsilon_x$, the beam energy and the energy spread $\sigma_{E}$ are shown. Fig.~\ref{fig:phasespaces_combined_inplasma_150MeV}(a) shows the simulated transverse and longitudinal phase spaces after the beam has been accelerated through the LPA. It is seen that we obtain an emittance of around 0.5 $\mu$m with a final beam energy of 1 GeV. A three-sigma cut was applied in the determination of the various observables, leading to 1\% of the total number of particles being cut (particles far out in the tail would otherwise bias the calculated observables). 

Achieving emittance preservation and a small energy spread is essential for a usable beam quality from a LPA. The slice energy spread for the beam after the plasma cell is very good and amounts to 0.3\% in our case which is still on the high side compared to facilities like FLASH \cite{Flash}. However, for applications such as FEL, large energy spread can be tolerated by reducing the slice energy spread by means of bunch decompression as demonstrated in \cite{Maier:2012LP_FEL}. Recently, a scheme has been proposed to diminish the energy spread by using a chicane between two plasma stages \cite{Pousa:PRL2019Multistagedechirping}.  

A perturbed beam, according to mismatch case 1 of Fig.~\ref{fig:Histograms150}, was also simulated through the plasma cell. In this mismatch case, the offsets were introduced in the beam parameters at the exit of BC as well as in the quadrupole triplet. Fig.~\ref{fig:beamevolution_inplasma_150MeV} shows the beam evolution of the mismatched beam in comparison to the matched case. In both cases, the slice emittance is still below 0.5 µm as shown in Fig.~\ref{fig:phasespaces_combined_inplasma_150MeV}. From Fig.~\ref{fig:beamevolution_inplasma_150MeV} and \ref{fig:phasespaces_combined_inplasma_150MeV}, it is seen that the proposed tolerances in the mismatch case 1 are sufficient to match the electron beam coming from the chicane to the plasma cell and to transport it with somewhat deteriorated but still good quality through a LPA.

\section{Conclusion}

The design for a final focus system, an experimental beamline and a LPA with external injection has been developed for the ARES linac in the SINBAD facility at DESY. Detailed numerical simulations show that the electron beam can be transported and focused to a plasma cell. The studies include the effects of space charge. The electron beam has adequate transverse symmetry and is well matched into a plasma channel with plasma ramps. The longitudinal phase space is preserved with a 1 kA peak current, as approaching the requirements for several use cases. We have performed a sensitivity analysis of the PMQ triplet for understanding the tolerances and to mitigate the effect of imperfections of the final focus system. The performed error analysis, which is specific to the system under study but could be generalised for any quadrupole triplet, gives a useful estimate about the performance of the final focus system and suggests critical parameters in the implementation of the experiment. A plasma simulation shows that external injection of short electron bunches into a LPA at ARES can achieve high beam quality and can constitute a stepping stone towards a staged multi GeV high performance plasma accelerator.

\begin{acknowledgments}
The authors acknowledge valuable discussions with K. Floettmann, D. Marx, F. Mayet and F. Jafarinia. Special thanks to W. Leemans for his support and discussion. The authors would also like to acknowledge the FBPIC developers and contributors.
\end{acknowledgments}
\bibliography{References}
\end{document}